\documentclass[preprint,12pt]{elsarticle}

\usepackage{graphicx}
\usepackage{amssymb}
\usepackage{lineno}
\usepackage{url}

\journal{Communications in Nonlinear Science and Numerical Simulation}

\begin{document}

\begin{frontmatter}

\title{On the uncertainty of real-time predictions of epidemic growths: a COVID-19 case study for China and Italy}

\author[aff1]{Tommaso Alberti}
\address[aff1]{INAF - Istituto di Astrofisica e Planetologia Spaziali, via del Fosso del Cavaliere 100, 00133 Roma, Italy}

\author[aff2,aff3,aff4]{Davide Faranda}
\address[aff2]{Laboratoire des Sciences du Climat et de l'Environnement, 5 CEA Saclay l'Orme des Merisiers, UMR 8212 CEA-CNRS-UVSQ, 6 Universit\'e Paris-Saclay \& IPSL, 91191 Gif-sur-Yvette, France}
\address[aff3]{London Mathematical Laboratory, 8 Margravine Gardens, London, W6 8RH, UK}
\address[aff4]{LMD/IPSL, Ecole Normale Superieure, 9 PSL research University, Paris, France}

\begin{abstract}
While COVID-19 is rapidly propagating around the globe, the need for providing real-time forecasts of the epidemics pushes fits of dynamical and statistical models to available data beyond their capabilities. Here we focus on statistical predictions of COVID-19 infections performed by fitting asymptotic distributions to actual data. By taking as a case-study the epidemic evolution of total COVID-19 infections in Chinese provinces and Italian regions, we find that predictions are characterized by large uncertainties at the early stages of the epidemic growth. Those uncertainties significantly reduce after the epidemics peak is reached. Differences in the uncertainty of the forecasts at a regional level can be used to highlight the delay in the spread of the virus. Our results warn that long term extrapolation of epidemics counts must be handled with extreme care as they crucially depend not only on the quality of data, but also on the stage of the epidemics, due to the intrinsically non-linear nature of the underlying dynamics. These results suggest that real-time epidemiological projections should include wide uncertainty ranges and urge for the needs of compiling high-quality datasets of infections counts, including asymptomatic patients.
\end{abstract}


\end{frontmatter}

%\linenumbers

\section{Introduction}

The COVID-19, a disease caused by the SARS-CoV-2 virus, was firstly reported in the Hubei province on 31 December 2019 when the WHO China Country Office was informed of cases of pneumonia unknown etiology detected in Wuhan City \cite{Service20,Huang20,Guan20}. On 7 January 2020 the Chinese authorities identified this virus as a zoonotic virus belonging to the family of coronavirus \cite{WHO05,Li20,Lipsitch20}. Its diffusion rapidly spread over all Chinese provinces and nearest countries (Thailand, Japan, Korea) \cite{Zhu20}. On 23 January, although still unknown the initial source of the epidemic, the evidence that 2019-nCoV spreads from human-to-human and also across generations of cases quickly increases \cite{Wang20,Phan20}. On 30 January, the World Health Organization (WHO) declared the outbreak to be a public health emergency of international concern \cite{WHO30}, believing that it is still possible to interrupt the virus spread by putting in place strong measures for early detecting, isolating, and treating cases, for tracing back all contacts, and for promoting social distancing measures \cite{WHO30,Parmet20,Haffajee20}. The main driver of transmission is still an open question \cite{Flaxman20,Rothe20}, and preliminary estimates of the median incubation period are 5-6 days (ranging between 2 and 14 days) \cite{Lauer20}. On 21 February a cluster of cases was detected in Italy (Lombardia), then on 23 February 11 municipalities in northern Italy were identified as the two main Italian clusters and placed under quarantine \cite{DPCM2302}, on 9 March the quarantine has been expanded to all of Italy \cite{DPCM0903}, on 11 March all commercial activity except for supermarkets and pharmacies were prohibited \cite{DPCM1103}, and on 22 March all non-essential businesses and industries were closed \cite{DPCM2203} and additional restrictions to movement of people were introduced \cite{Rosenbaum20,Chinazzi20}. 

Meanwhile, the quarantined Chinese regions observed a fast decrease in the number of cases in Hubei and a moderate decrease in other affected regions, at the same time the virus internationally spread, and on 11 March the WHO declared COVID-19 a pandemic \cite{Fauci20,WHO11}. To date, there are more than 1 million confirmed cases over the globe, more than 60000 deaths, and the most affected areas are the European region and the United States. While three months were needed to reach the first 100000 confirmed cases, only 23 days were sufficient to multiply by eight the counts, a typical signature of the exponential spreading of viruses. The reason for such high infectivity  are currently being explored in clinical studies and numerical simulations~\cite{Spinello2020}. Due to the fast spread of the virus and the severity of symptoms, restrictive confinement measures have been imposed in many countries. They were based on asymptotic extrapolation of infection counts obtained on the basis of compartmental epidemic models as the Susceptible-Exposed-Infected-Recovered (SEIR) model and their variants \cite{Brauer08} or on agent-based models~\cite{chang2020modelling}. Unfortunately, predictions made using these models are extremely sensitive to the underlying parameters and the quality of their extrapolation is deeply affected from both the lack of high-quality datasets as well as from the intrinsic sensitivity of the dynamics to initial conditions in the growing phase~\cite{Faranda20}. Moreover, in order to provide reliable estimates of asymptotic infection counts, a knowledge of asymptomatic populations is needed. These data are currently almost unavailable and affected by great uncertainties.

Another possibility is to extrapolate the number of infections by means of fitting asymptotic distributions to actual data. Using these phenomenological statistical approach, we compare the behavior of epidemic evolution across China and Italy. The assumption beyond those fits is that typical curves of total infections in SEIR models display a sigmoid shape~\cite{wilcosky1985comparison}. Sigmoid functions such as the logistic or Gompertz  can therefore be used to fit actual data. When data are collected with the same protocols, e.g., in China and Italy, where tests are performed only to symptomatic patients,  the statistical fitting can therefore provide an extrapolation of how many symptomatic cases should be recorded, although it will not inform about the real percentage of infected population~\cite{batista2020estimation}. We found that predictions are characterized by large uncertainties at the early stages of the epidemic growth, significantly reducing when a mature stage or a peak of infections are reached. This is observed both in China and in Italy, although some differences are observed across the Italian territory, possibly related with the time delayed diffusion of epidemic into the different Italian regions. Finally, we also estimate infection increments for each Italian region, with being the uncertainty significantly reduced for Northern and Central regions, while a larger one is found for Southern regions. These results can be helpful for each epidemic diffusion, thus highlighting that the confinement measures are fundamental and more effective in the early stages of the epidemic evolution (the first 7 days), thus producing a different spread across provinces/regions as these measures are considered. 

\section{Data}

\begin{figure}[h]
\centering\includegraphics[width=\linewidth]{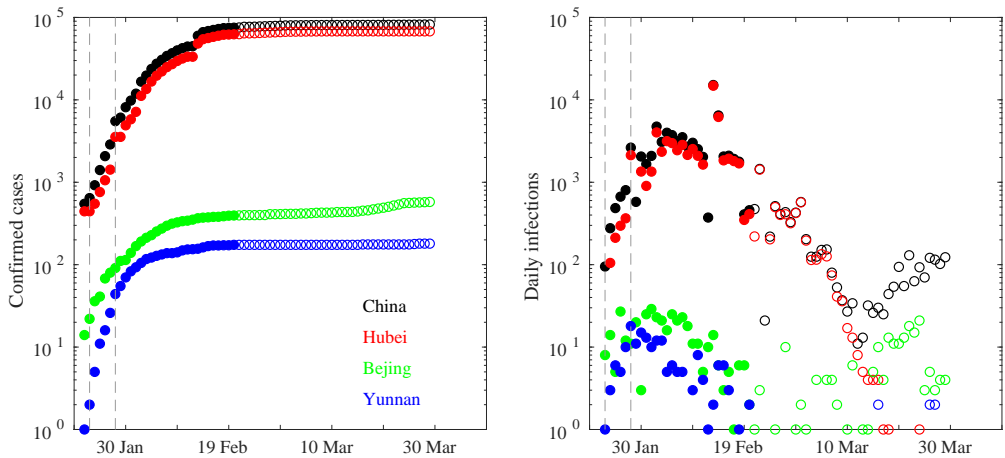}
\caption{The total number of confirmed infections (left panel) and the daily infections (right panel) for China and three Chinese provinces of Bejing, Hubei, and Yunnan. Filled circles refer to the first 30 days of the epidemic diffusion. The vertical dashed lines mark the times when Chinese government applied lock-down restrictions on 23 January and 28 January, respectively.}
\label{fig1}
\end{figure}

Data for the Chinese provinces are obtained from the data repository for the 2019 Novel Coronavirus Visual Dashboard operated by the Johns Hopkins University Center for Systems Science and Engineering (JHU CSSE), freely available at \url{https://github.com/CSSEGISandData/COVID-19}. Fig.~\ref{fig1} reports the total number of confirmed infections (left panel), thus including actual positive people to COVID-19, recovered and deaths for China and three Chinese provinces of Bejing, Hubei, and Yunnan, and the daily infections (right panel), during the period between 22 January and 30 March.  

\begin{figure}[h]
\centering\includegraphics[width=\linewidth]{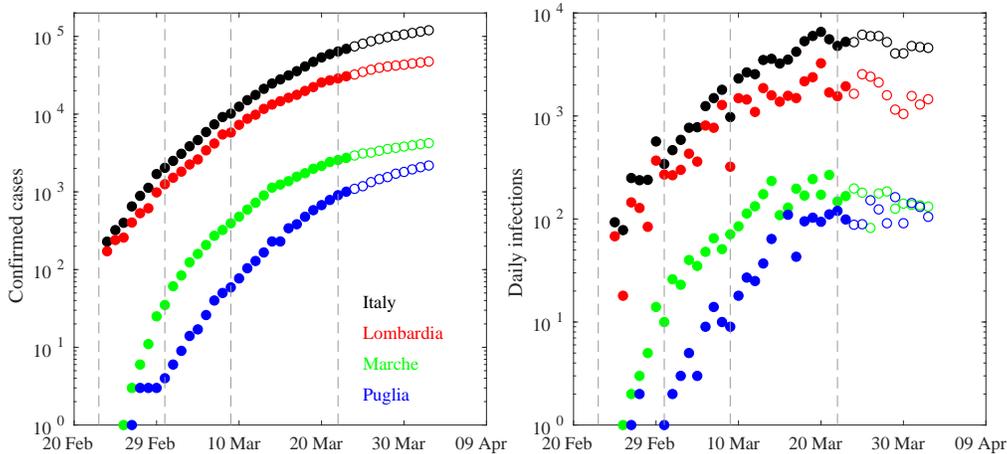}
\caption{The total number of confirmed infections (left panel) and the daily infections (right panel) for Italy and three Italian regions of Lombardia, Marche, and Puglia. Filled circles refer to the first 30 days of the epidemic diffusion. The vertical dashed lines mark the times when the Italian government applied lock-down restrictions on 23 February, 01 March, 09 March, and 22 March, respectively.}
\label{fig2}
\end{figure}

Data for the Italian regions are instead derived from the repository freely available at \url{https://github.com/pcm-dpc/COVID-19} where data are collected from the Italian Protezione Civile from 24 February 2020. Data used here were last downloaded on 02 April, thus covering the period 24 February - 02 April, as shown in Fig.~\ref{fig2}. 
It is evident that although the increments of infections started about 1 month after the Chinese epidemic Italy has fast reached and exceeded the Chinese peak values of $\sim$80000 infections. Moreover, it is also apparent that epidemic diffusion in China reached its peak within $\sim$20 days from the first restriction operated to the Hubei region on 23 January. Conversely, the Italian restrictions seem to become more efficient only when the Italian government adopted a lock-down confinement on 9 March \cite{DPCM0903}. 

\section{Methods}

A data-driven way to extrapolate future phases of an epidemic growth \cite{Li12,Kumar20,Ma20} is to use a generalized logistic distribution for fitting the total cumulative number of infections as \cite{Gompertz85,Chowell17,Burger19}
\begin{equation}
\label{eq:logistic}
    C(t) = \frac{\alpha}{1 + \beta \, e^{-\gamma t}}
\end{equation}
being $\alpha$, $\beta$, and $\gamma$ the parameters of the model. They can be fitted, e.g., using  Nonlinear least-squares solver, with the Levenberg-Marquardt algorithm and the bisquare weight methods to minimize a weighted sum of squares. Here we use a MATLAB function to perform the fits. As recently pointed out in \cite{Faranda20} in the early stages of the epidemics, the smoothness of COVID-19 cumulative infections data could lead to very uncertain predictions although with very good R$^2$. To avoid this, here we focus only on Chinese and Italian data, that, to date, represent a  mature stage of the epidemics. This implies, as we will show, that the significance of the logistic fit can be assigned with greater confidence \cite{Faranda20}. 
We remark however, that when confinement measures are applied, the basic reproduction number $R_0$, which regulates the growth of infections, is reduced \cite{Petrosillo20}. We are therefore in presence not of a single logistic distribution, but of a mixture of distributions with control parameters changing in time as different phases of epidemic diffusion are reached. Confinement measures can reduce $R_0$ from the exponential-like behavior of an uncontrolled growing phase, to a smoother logistic growth phase. Our goal here is to use the a-priori knowledge of the introduction of confinement measurements to investigate the perfomrance of statistical prediction of infection counts for different epidemic phases. Thus, we perform logistic fits as in Eq.~(\ref{eq:logistic}) in the following time intervals:
\begin{itemize}
    \item the first 30 days of epidemic growth, as reported in Figs.~\ref{fig3}-\ref{fig4} by black lines, thus to consider how restrictions measure globally affect the diffusion;
    \item the first 7 days, roughly corresponding to the time interval during which first restriction measures are adopted both in China and Italy, although not still completely efficient (red lines in Figs.~\ref{fig3}-\ref{fig4});
    \item the first 14 days, corresponding to the time interval in which the initial confinement measures should lead the first effects (blue lines in Figs.~\ref{fig3}-\ref{fig4});
    \item the time interval between the 8$^{th}$ and the 14$^{th}$ day to investigate how the epidemic would be grown if starting from initial restrictions (green lines in Figs.~\ref{fig3}-\ref{fig4});
    \item the time interval between the 15$^{th}$ and the 30$^{th}$ day to investigate the efficiency of restriction measures (magenta lines in Figs.~\ref{fig3}-\ref{fig4}).
    \end{itemize}
In this way we can investigate both the efficiency of restriction measures in containing epidemic growth as well as the stability of prediction models based on logistic distribution fitting procedures. Moreover, to assess the significance of fits we assume that the last point of the fitting range could be affected by a $\pm$30\% error. This allows us to provide a simple way to estimate confidence intervals for our fits \cite{Faranda20}. 
Finally, the Kolmogorov-Smirnov (K-S) test \cite{Kolmogorov33,Smirnov48,Stephens74} is also used to obtain a test decision for the null hypothesis that the observed data are from the same logistic distribution as derived from the logistic fits under the different time intervals. This allows to test the efficiency in delivering reliable forecasts at different stages of the epidemic growth. The test is based on evaluating the maximum distance between the empirical distribution functions coming from two different samples $x_{1,n}$ and $x_{2, m}$, being $n$ and $m$ the length of samples. By defining the Kolmogorov-Smirnov statistic as
\begin{equation}
    D_{n, m} = \sup_ x \left| F_{1, n}(x) - F_{2, m}(x) \right|,
\end{equation}
where $F_{1, n}(x)$ and $F_{2, m}(x)$ are the empirical distribution functions of the two samples, respectively, the null hypothesis is rejected at the confidence level $\alpha$ if
\begin{equation}
    D_{n, m} > c(\alpha) \sqrt{\frac{n + m}{n \cdot m}}.
\end{equation}
When $m = n$ a general relation can be found for $D_{n}(\alpha)$ as
\begin{equation}
    D_n(\alpha) > \frac{1}{\sqrt{n}} \sqrt{- \log\left(\frac{\alpha}{2}\right)}.
    \label{eq:Dn}
\end{equation}
The value of $c(\alpha)$ for the most common levels of $\alpha$ are reported in Table \ref{tab:c}. 
\begin{table}[h]
    \centering
  \begin{tabular}{|c|ccccc|}
        \hline
        $\alpha$    & 0.20  & 0.15  & 0.10  & 0.05  & 0.01 \\
        $c(\alpha)$ & 1.073 & 1.138 & 1.224 & 1.358 & 1.628 \\
        \hline
    \end{tabular}
    \caption{The value of $c(\alpha)$ for the most common levels of $\alpha$.}
    \label{tab:c}
\end{table}

The closer the observed statistics $D_{n, obs}$ is to 0 the more likely it is that the two samples were drawn from the same distribution with being $D_{n, obs} < D_n(\alpha)$. The use of the K-S test has two main advantages: i) the distribution of the K-S test statistic itself does not depend on the underlying cumulative distribution function being tested, and ii) it is an exact test \cite{Anderson54,Marsaglia03,Encyclopedia08,Marozzi13}. Moreover, it is specifically designed for testing if data come from a normal, a log-normal, a Weibull, an exponential, or a logistic distribution \cite{Marsaglia03,Massey52}. Thus, it is particularly suitable for our investigations, being also a non-parametric and robust technique since it is not based on strong distributional assumptions \cite{Marsaglia03,Massey52,Ghidey10,Marozzi13}.  

\section{Epidemic diffusion through Chinese provinces} 

\begin{figure}[h]
\centering\includegraphics[width=0.8\linewidth,height=0.7\paperheight]{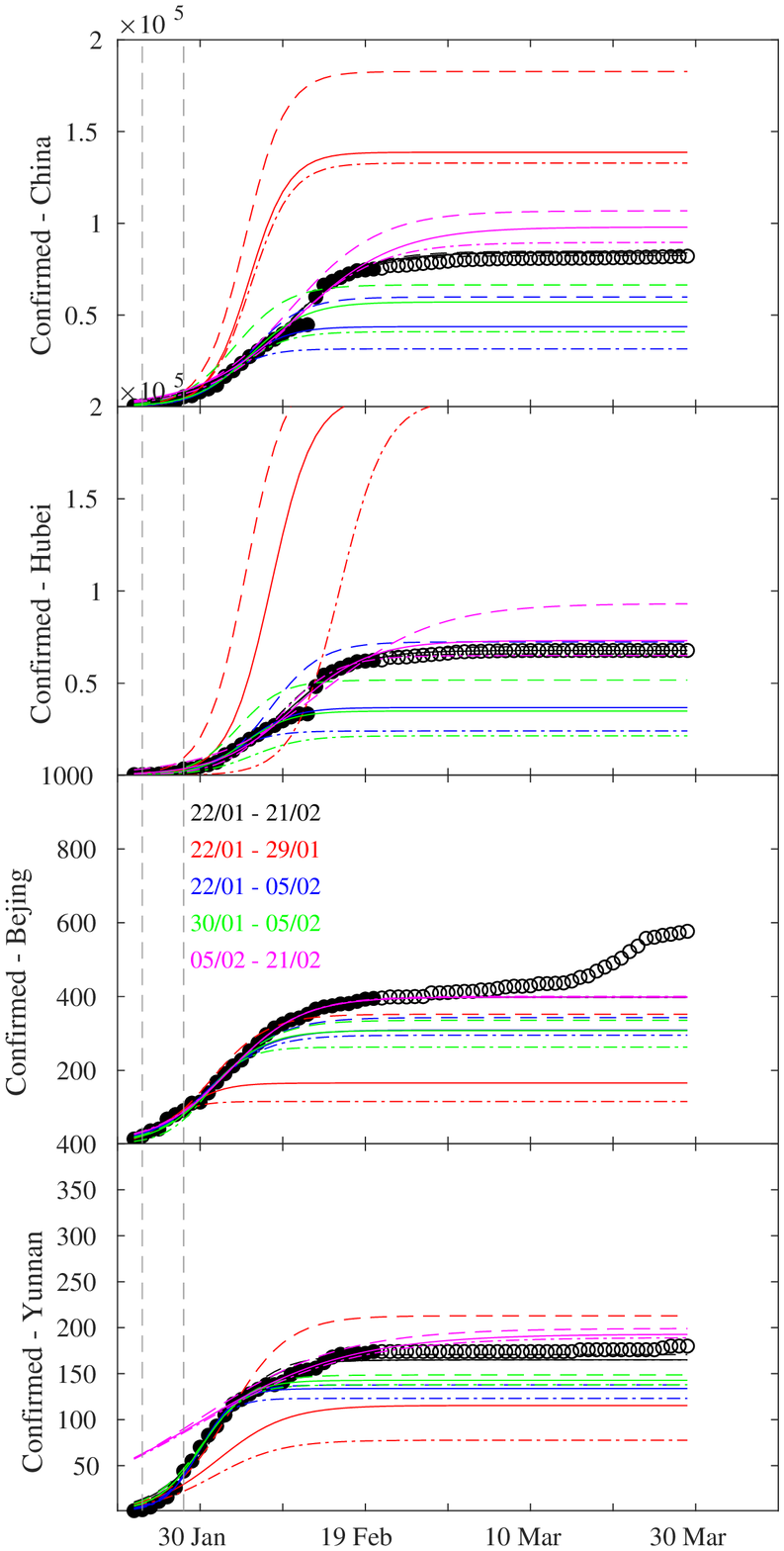}
\caption{Logistic fits during the different time intervals of epidemic across Chinese provinces, together with the confidence lines. From top to bottom: China and three provinces (Bejing, Hubei, Yunnan). The vertical dashed lines mark the times when Chinese government applied lock-down restrictions on 23 January and 28 January, respectively.}
\label{fig3}
\end{figure}

Fig.~\ref{fig3} shows logistic fits for different phases of epidemic across Chinese provinces, together with  upper and lower confidence bounds, obtained as outlined in the previous section.
Early stage of epidemic propagation is  characterized by a larger confidence interval (red lines in Fig.~\ref{fig3}), thus highlighting the difficulty in making early reliable predictions of epidemic growth, with an exponential-like behavior. The confidence interval becomes narrower as the growth rate reduces, as for the case of the provinces of Bejing and Yunnan being less affected from COVID-19 infections with respect to the Hubei, the latter mostly contributing to the overall epidemic growth in China. The logistic fit becomes more stable, being characterized by a narrower estimates of confidence intervals, when the first two weeks are considered (blue lines in Fig.~\ref{fig3}), possibly related to the initial efficiency of restriction measures. This could be also due to both the limited number of points of the fitting range as well as to the particular phase of the epidemic growth. However, by comparing the confidence intervals of logistic fits performed using the first week (22/01 - 29/01, red lines in Fig.~\ref{fig3}) and the second week (30/01 - 05/02, green lines in Fig.~\ref{fig3}) it is possible to note that the stability increases for this second interval for all Chinese provinces, thus suggesting that estimates are significantly dependent on the particular epidemic phase considered. Indeed, the stability significantly increases when the logistic fit is performed on time intervals that do not include the first week of the epidemic growth (green and magenta lines in Fig.~\ref{fig3}), suggesting that credible predictions could be assigned with a large confidence by means of a logistic fit if the beginning of the outbreak is not considered. However, the narrowest estimates of significance levels is obtained when the first 30 days are considered, thus also including the beginning of the outbreak, possibly suggesting that fits become more and more stable if data are collected at a mature stage of the epidemic growth. This is clearly visible for all Chinese provinces, apart for the slight discrepancy observed for the Bejing province where some returned cases from outside China were observed from 20 March. Finally, we assess the statistical discrepancy of the logistic fits from the observed data by performing the Kolmogorov-Smirnov (K-S) test those results for the 95\% confidence level are reported in Table \ref{tab:KSchina}.
\begin{table}[h]
  \centering
  \begin{tabular}{|c|cccc|}
        \hline
                        & \multicolumn{4}{c|}{$D_{n, obs}$} \\
        \hline
          Time interval & China     & Hubei     & Bejing    & Yunnan    \\
        \hline
          22/01 - 29/01 & 0.750     & 0.750     & 0.750     & 0.625     \\
          22/01 - 05/02 & 0.500     & 0.475     & 0.550     & 0.450     \\
          30/01 - 05/02 & 0.575     & 0.575     & 0.550     & 0.525     \\
          05/02 - 21/02 & {\bf 0.225}     & {\bf 0.150}     & {\bf 0.125}     & {\bf 0.125}     \\
          22/01 - 21/02 & {\bf 0.100}     & {\bf 0.100}     & {\bf 0.100}     & {\bf 0.100}     \\
        \hline
    \end{tabular}
    \caption{Results of the Kolmogorov-Smirnov test for the 95\% confidence level for the Chinese provinces. The decision to reject the null hypothesis is based on comparing the observed statistics $D_{n, obs}$ with the theoretical value $D_{n, th} = 0.2329$ obtained for the significance level $\alpha = 0.05$ as in Eq.~\ref{eq:Dn}. If $D_{n, obs} < D_{n, th}$ then the samples come from the same logistic distribution and corresponding values are reported in bold.}
    \label{tab:KSchina}
\end{table}

It can be noted that the statistical results obtained through the K-S test suggest that the fits performed by considering the time intervals from 22 January to 21 February as well as from 05 February to 21 February are statistically significant for reproducing the behavior of the observed number of infections at the 95\% significance level. This seems to support the that reliable predictions can be assessed only when a mature stage of the epidemic growth is approached/reached, while low-significant predictions can be released at the early stages of the epidemic diffusion.

\section{Epidemic diffusion through Italian regions} \label{sec:italy}
\begin{figure}[h]
\centering\includegraphics[width=0.8\linewidth,height=0.7\paperheight]{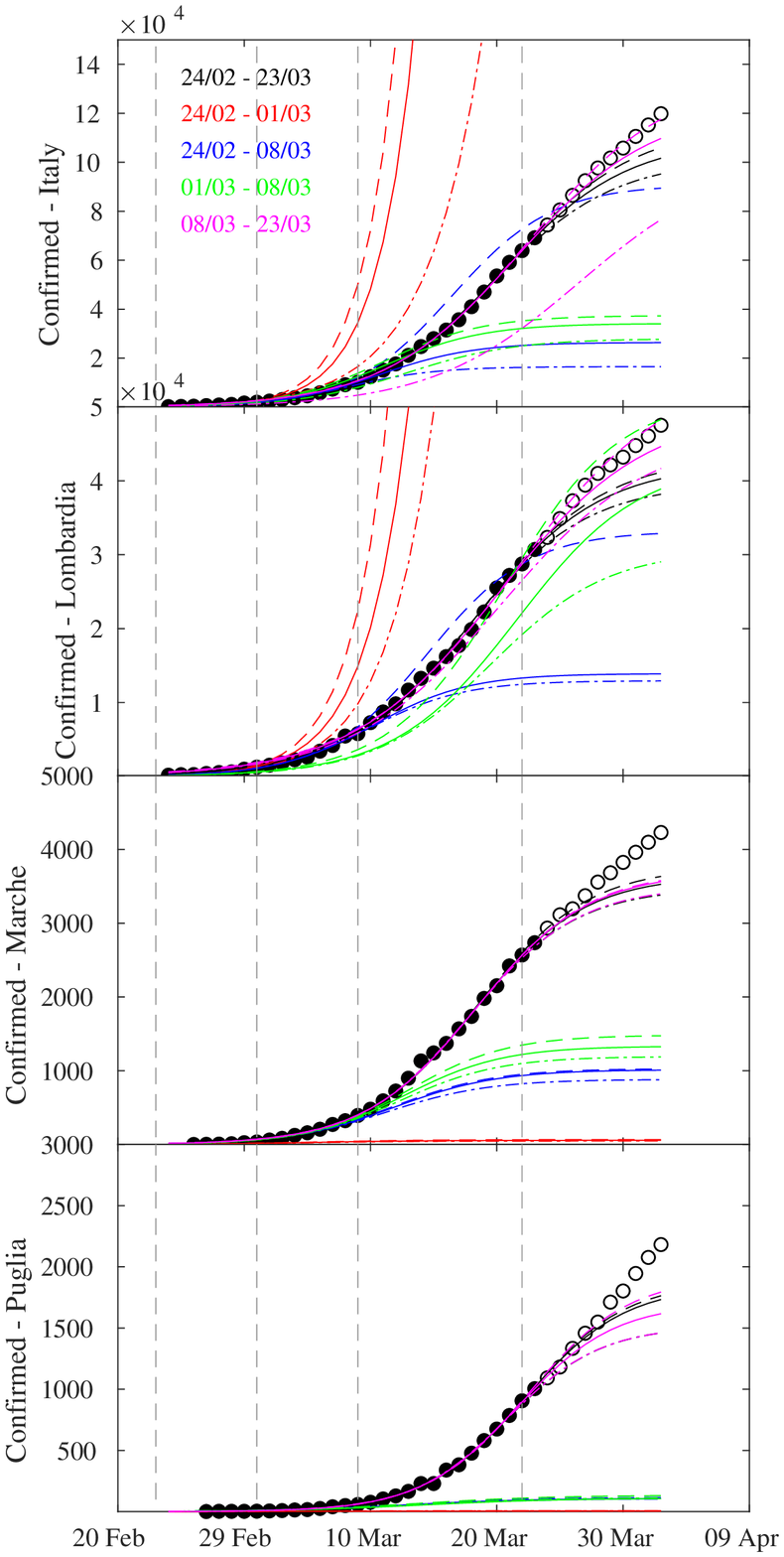}
\caption{Logistic fits during the different time intervals of epidemic across Italian regions, together with the confidence lines. From top to bottom: Italy and three regions (Lombardia, Marche, Puglia). The vertical dashed lines mark the times when Italian government applied lock-down restrictions on 23 February, 01 March, 09 March, and 22 March, respectively.}
\label{fig4}
\end{figure}

Fig.~\ref{fig4} shows logistic fits for different phases of epidemic across Italian regions, together with the upper and lower confidence lines.
As for Chinese provinces the early stage of epidemic diffusion is characterized by a larger confidence interval (red lines in Fig.~\ref{fig4}), again suggesting that reliable predictions of epidemic growth are particularly difficult in its early stages. Indeed, an exponential-like behavior is found for both the Italian territory and Lombardia, the latter being the first Italian region characterized by COVID-19 infections. As for China, confidence intervals become narrower as the growth rate reduces (see for example Marche or Puglia with respect to Lombardia), with the logistic fits also becoming more stable when the initial stages of the outbreak are removed (green and magenta lines in Fig.~\ref{fig4}). Unlike for Chinese regions, Italian regions present a wide range of different epidemic behaviors, that we investigate separately in the following.

\subsection{Epidemics growth in Lombardia}

As discussed above the initial phase is characterized by larger uncertainties and by an exponential-like behavior (red lines in Fig.~\ref{fig4}), thus suggesting a clear difficulty in making predictions of the growth in its early stage. When the first two weeks (e.g., 24/02 - 08/03) are considered (blue lines in Fig.~\ref{fig4}) a larger uncertainty is found, especially for the upper-bound confidence level. This underline the difficulty in making reliable estimates of its evolution. Similarly, the logistic fits performed between 01 March and 08 March (green lines in Fig.~\ref{fig4}) suggest that the first two weeks were particularly critical in Lombardia, while logistic fits become more stable when removing the beginning of the outbreak, leading to more reliable estimates of the epidemic growth (magenta lines in Fig.~\ref{fig4}). Finally, significance levels become narrower when the first 30 days are considered (e.g., 24/02 - 23/03), thus also including the beginning of the outbreak, possibly again suggesting that including data from the mature stage of the epidemic growth could allow to obtain more stable fits. We remark that, no matter the approach followed, logistic fits struggle to predict the number of infections of the successive days. This failure of statistical real-time forecasts of the epidemics could be related to all those factors that can change the instantaneous value of $R_0$, e.g., extended violations of the restriction measures, changes in testing protocols or delay in data reporting, changes in the virus characteristics. It is worthwhile to note that the above features are found for all Northern regions firstly affected from COVID-19 diffusion (not shown here). 

\subsection{Epidemics growth in Marche}

The epidemic growth throughout Marche, as well as throughout other Central regions (not shown), is different from Northern regions. Indeed, the first 7 days (e.g., 24/02 - 01/03, red lines in Fig.~\ref{fig4}) were not characterized by an exponential increase of infections, as the diffusion of the virus was pretty slow: logistic fits are therefore meaningless in this context. The exponential phase started in the second week, as we can see by fitting the first two week of the infection counts (e.g., 24/02 - 08/03, blue lines in Fig.~\ref{fig4}) or just the second week (e.g., from 01 March to 08 March, green lines in Fig.~\ref{fig4}). During this week, the number of infections significantly increases (272 confirmed cases) enabling better fits of data to logistic distributions. This suggests a time delayed propagation between Northern and Central regions. Indeed, the logistic fits become more stable, with narrower estimates of confidence intervals, when the time interval from 08 March to 23 March (magenta lines in Fig.~\ref{fig4}) or the first 30 days (e.g., 24/02 - 23/03, black lines in Fig.~\ref{fig4}) are taken into account, suggesting that credible predictions could be assigned with a large confidence when a mature stage of the epidemic growth is approached. However, as for Norther regions the logistic fits struggle to predict the number of infections of the successive days (i.e., after the first 30 days).

\subsection{Epidemics growth in Puglia}

A completely different scenario is found for Puglia and Southern regions (not shown). Logistic fits cannot be performed during during the first two weeks (e.g., from 24 February to 08 March), as the infection counts was not yet exponential. By considering the time interval between 08 and 23 March (magenta lines in Fig.~\ref{fig4}) and the first 30 days (e.g., 24/02 - 23/03, black lines in Fig.~\ref{fig4}) an increase in the confidence of logistic fits is found, although they struggle to predict the number of infections of the successive days (i.e., after the first 30 days). This is possibly due to the time delayed propagation of epidemic throughout Southern regions for which a mature stage is, to date, not yet reached. To support this hypothesis and to assess the statistical discrepancy of the logistic fits from the observed data we perform the Kolmogorov-Smirnov (K-S) test those results for the 95\% confidence level are reported in Table \ref{tab:KSitaly}.
\begin{table}[h]
  \centering
  \begin{tabular}{|c|cccc|}
        \hline
                        & \multicolumn{4}{c|}{$D_{n, obs}$} \\
        \hline
          Time interval & Italy     & Lombardia   & Marche  & Puglia  \\
        \hline
          24/02 - 01/03 & 0.825     & 0.800       & 0.800   & 0.800   \\
          24/02 - 08/03 & 0.575     & 0.550       & 0.650   & 0.800   \\
          01/03 - 08/03 & 0.550     & 0.425       & 0.600   & 0.800   \\
          08/03 - 23/03 & 0.325     & 0.325       & 0.400   & 0.400   \\
          24/02 - 23/03 & 0.350     & 0.325       & 0.400   & 0.400   \\
        \hline
    \end{tabular}
    \caption{Results of the Kolmogorov-Smirnov test for the 95\% confidence level for the Italian regions. The decision to reject the null hypothesis is based on comparing the observed statistics $D_{n, obs}$ with the theoretical value $D_{n, th} = 0.3037$ obtained for the significance level $\alpha = 0.05$ as in Eq.~\ref{eq:Dn}. If $D_{n, obs} < D_{n, th}$ then the samples come from the same logistic distribution and corresponding values are reported in bold.}
    \label{tab:KSitaly}
\end{table}

It is interesting to note that, although lower values of $D_{n, obs}$ are observed when a more mature stage of the epidemic growth is considered in the fitting range, as for example for time intervals from 24 February to 23 March as well as from 08 to 23 March, the observed values $D_{n, obs}$ are all above the statistical threshold of $D_{n, th} = 0.3037$. This suggests that a mature stage is, to the date of 23 March, not yet reached, although Northern and Central regions are characterized by lower values than the Southern ones, thus possibly related to the time delayed propagation of epidemic throughout Southern regions.

\section{Estimation of infections for Italy and their peak time}

\begin{figure}[h]
\centering\includegraphics[width=0.8\linewidth,height=0.7\paperheight]{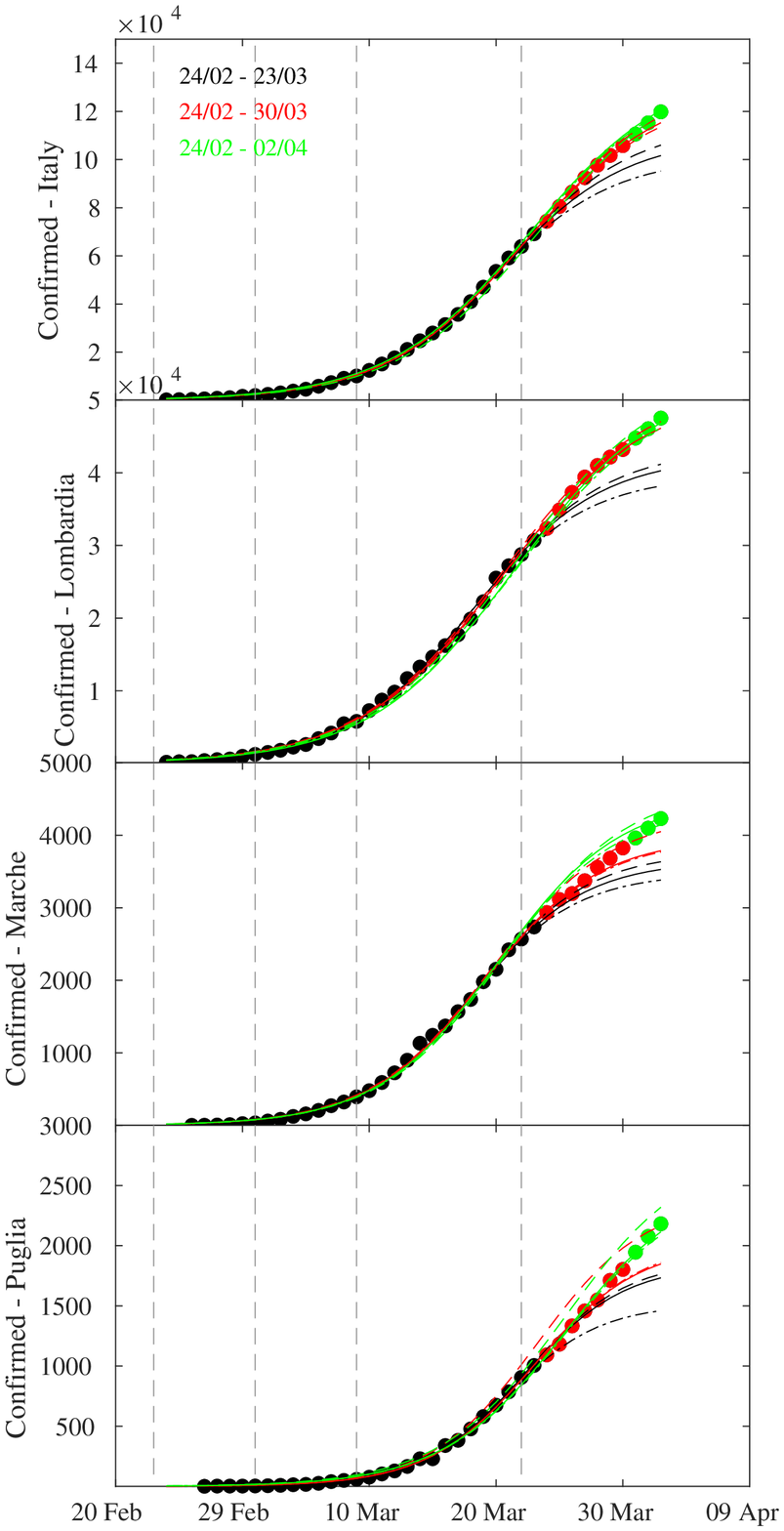}
\caption{Logistic fits during the different time intervals of epidemic across Italian regions, together with the confidence lines. From top to bottom: Italy and three regions (Lombardia, Marche, Puglia). The vertical dashed lines mark the times when Italian government applied lock-down restrictions on 23 February, 01 March, 09 March, and 22 March, respectively.}
\label{fig5}
\end{figure}

As discussed in Section \ref{sec:italy} all performed logistic fits struggle to predict the number of infections of the successive days (i.e., after the first 30 days), thus we performed and compare logistic fits in three time intervals: (i) the first 30 days (e.g., from 24 February to 23 March), (ii) the first 37 days (e.g., from 24 February to 30 March), and (iii) the overall period from 24 February to 02 April. The results of the Kolmogorov-Smirnov test for the 95\% confidence level are reported in Table \ref{tab:KS}, while the behavior of logistic fits are shown in Fig.~\ref{fig5}.
\begin{table}[h]
  \centering
  \begin{tabular}{|c|cccc|}
        \hline
                        & \multicolumn{4}{c|}{$D_{n, obs}$} \\
        \hline
          Time interval & Italy   & Lombardia     & Marche    & Puglia  \\
        \hline
          24/02 - 23/03 & 0.350   & 0.325         & 0.400     & 0.400   \\
          24/02 - 30/03 & {\bf 0.150} & {\bf 0.150} & {\bf 0.250}   & {\bf 0.275}       \\
          24/02 - 02/04 & {\bf 0.100} & {\bf 0.100} & {\bf 0.175}   & {\bf 0.200}     \\
    \hline
    \end{tabular}
    \caption{Results of the Kolmogorov-Smirnov test for the 95\% confidence level for the Italian regions. The decision to reject the null hypothesis is based on comparing the observed statistics $D_{n, obs}$ with the theoretical value $D_{n, th} = 0.3037$ obtained for the significance level $\alpha = 0.05$. If $D_{n, obs} < D_{n, th}$ then the samples come from the same logistic distribution.}
    \label{tab:KS}
\end{table}

\begin{figure}[h]
\centering\includegraphics[width=0.8\linewidth,height=0.7\paperheight]{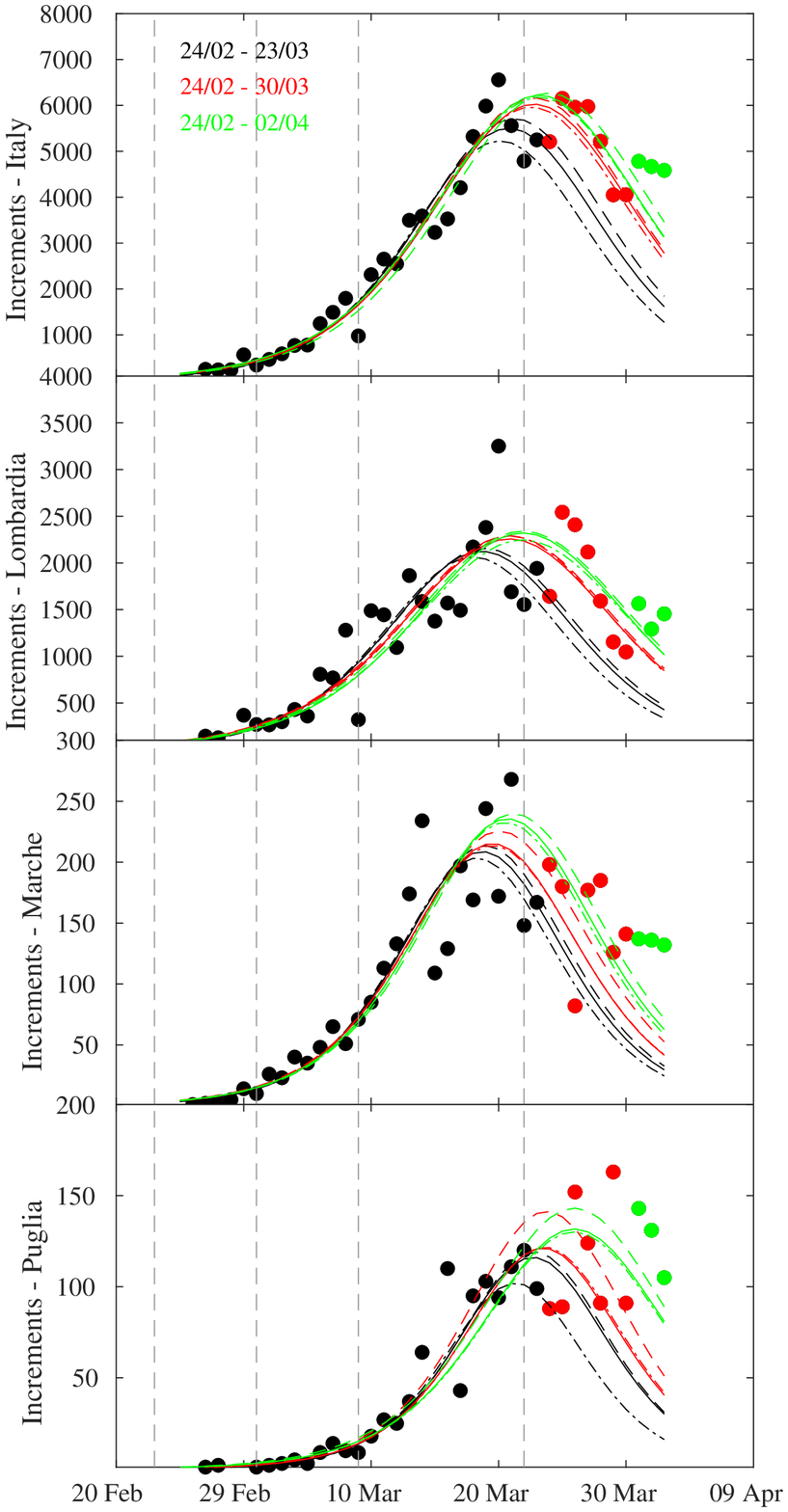}
\caption{Estimation of daily infections and their peak time during three different time intervals of epidemic across Italian regions, together with the confidence lines. From top to bottom: Italy and three regions (Lombardia, Marche, Puglia). The vertical dashed lines mark the times when Italian government applied lock-down restrictions on 23 February, 01 March, 09 March, and 22 March, respectively.}
\label{fig6}
\end{figure}

It is interesting to note that all regions and Italy are characterized by lower values of $D_{n, obs}$, below the theoretical value $D_{n, th} = 0.3037$,  when including the next 7 days (e.g., by considering the period between 24 February and 30 March) to the logistic fits and when considering the whole time range (e.g., 24/02 - 02/04). Lombardia presents lower values of the K-S statistics $D_{n, obs}$ than those for Marche and Puglia, together with a narrower confidence interval when including the successive days, not observed for both Marche and Puglia. Particularly for Puglia the confidence interval remains practically unchanged, thus suggesting that logistic fits are not still stable, possibly due to the fact that Southern regions have not yet reached a mature stage of the epidemic growth. This difference in terms of stability of logistic fits as well as on confidence of reliable estimates can be clearly seen by looking at the behavior of estimated daily increments. Days of peak significantly depends on the fitting range for Puglia, while the estimation of this quantity is more stable for Lombardia and Marche, as shown in Fig.~\ref{fig6}.
Indeed a wider discrepancy is found between daily increments and estimates for logistic fits performed during the three intervals, obviously affecting both the peak time estimation and its value. By comparing our estimates and data collected from the daily report of the Italian Protezione Civile (\url{https://github.com/pcm-dpc/COVID-19}) we found that the discrepancy significantly increases when moving from Northern to Southern regions, where it can also reach an error which is comparable with the predicted value. This could be the reflection of at least two different factors: i) the epidemic growth is in a more mature phase in the Northern and Central regions with respect to the Southern ones, where it began with a time delay ranging from 3 to 14 days, and ii) the higher ratio between the observed cases and the number of tests carried out for Southern regions with respect to the rest of Italy (see \url{https://github.com/pcm-dpc/COVID-19}). These two factors could affect the performance of the logistic fits for the Southern regions of Italy, being characterized by wider uncertainties with respect to the rest of Italy. Thus, our results suggest that estimates of the ending of epidemic growth are affected by the statistical uncertainties, by the delayed propagation of infections through the different regions, and by the effective respect of the guidelines in terms of confinement measures.

\section{Conclusion}

In this paper we investigated the behavior of predictions of COVID-19 infections on the particular phase of its growth and propagation in a specific country, as well as, on the effectiveness of social distancing and confinement measures. By analyzing the epidemic evolution in China and Italy we find that predictions are characterized by large uncertainties at the early stages of the epidemic growth, significantly reducing when  the epidemics peak is past, independently on how this is reached. While infection counts for different Chinese provinces show a synchronised behavior, counts for Italian regions point to  different epidemic phases. While the epidemic peak has been likely reached in the Northern and Central regions, COVID-19 infections are still in a growing phase for Southern regions, with a delay ranging from 3 to 14 days. By assessing the performance of logistic fits we assess that a wider uncertainty is found during the first week of epidemic propagation. Uncertainty is reduced when data from the very beginning of the breakout are removed from the datasets. Moreover, the estimated infection increments are extremely sensitive to the epidemic growth stage and to the last points considered to perform statistical extrapolations. Higher significance levels are reached for the more mature stages of the epidemic growth.

Our results aim at providing some guidelines for real-time epidemics forecasts which should be applicable to other viruses and outbreaks. Real-time forecasts of the epidemics are, to date, a societal need more than a scientific field. They are crucial to plan the duration of confinement measures and to define the needs for health-care facilities.  The aim of this letter was to show that those extrapolations crucially depend not only on the quality of data, but also on the stage of the epidemics, due to the intrinsically non-linear nature of the underlying dynamics. This prevents from performing successful long-term extrapolations of the infection counts with statistical models. On the other hand, dynamical models such as those based on compartments or agent dynamics, need to be initialized with quality data, faithfully representing the infected populations including asymptomatic patients \cite{Faranda20}. It is therefore crucial to pursue national health systems to provide the most transparent and extended datasets as possible and obtain high quality datasets to initialize those models. We remind that only dynamical models can provide a coherent representation and evolution of the epidemics, as they are effectively based on the conservation of the total number of individuals.

%\section*{Acknowledgments}

\bibliographystyle{elsarticle-num-names}
\bibliography{AlbertiandFaranda.bib}

\end{document}